# Photoconductive gain in semiconductor quantum wires

K.-D. Hof, C. Rossler, S. Manus, J.P. Kotthaus, A.W. Holleitner*, D. Schuh**, W. Wegscheider**

*Department für Physik and Center for Nanoscience, Ludwig-Maximilians Universität, Geschwister-Scholl-Platz 1, D-80539 München, Germany.*
*Walter Schottky Institut and Physik Department, Technische Universität München, D-85748 Garching, Germany*
**Institut für Experimentelle und Angewandte Physik, Universität Regensburg, D-93040 Regensburg, Germany*

**Abstract.** We report on a photoconductive gain in semiconductor quantum wires which are lithographically defined in an AlGaAs/GaAs quantum well via a shallow-etch technique. The effect allows resolving the one-dimensional subbands of the quantum wires as maxima in the photoresponse across the quantum wires. We interpret the results by optically induced holes in the valence band of the quantum well which shift the one-dimensional subbands of the quantum wire. Here we demonstrate that the effect persists up to a temperature of about 17 Kelvin.

## INTRODUCTION

Based on a proposal by Q. Hu in 1993, there have been several theoretical and experimental studies on the conductive photoresponse of semiconductor quantum wires (QWs) [1-9]. Far-infrared photons can give rise to thermopower [2-6] and rectification phenomena [7] across the QWs. Visible and near-infrared photons can result in a photoconductive gain across semiconductor QWs [8,9]. A persistent gain can be understood by the capacitive influence of optically excited holes being trapped at impurity states in close vicinity of the QW [8]. A dynamic photoconductive gain can be detected if optically excited electron-hole pairs are spatially separated because of the internal potential landscape of the QWs [9]. In the latter case, the one-dimensional subbands and the conductance of the QWs is capacitively influenced by the optically induced holes captured at the edges of the QWs [9,10]. The dynamic photoconductive gain is limited in time by the recombination lifetime of the spatially separated electrons and holes [9]. In this contribution, we demonstrate that the described dynamic photoconductive gain effect can be detected up to temperatures of about 17 K. At this temperature the thermal energy is comparable to the subband energy of the QWs, and in turn, fewer holes are captured at the edges of the QWs, and the recombination of the spatially separated electrons and holes is enhanced.

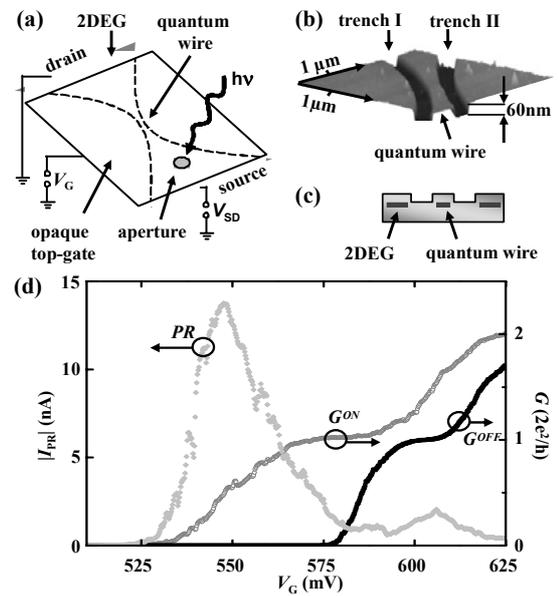

**FIGURE 1.** (a) A lateral constriction in a two-dimensional electron gas (2DEG) between source and drain contacts forms a semiconductor quantum wire (QW). An aperture in the opaque gate close to the constriction defines the position where the underlying 2DEG is optically excited [9]. (b) and (c): The QW is defined by two adjacent trenches I and II in the 2DEG via a shallow-etch technique. (d) Conductance and photoresponse data of a QW at $T = 2.2$ K, $f_{CHOP} = 313$ Hz and $U_{SD} = -1.5$ mV.

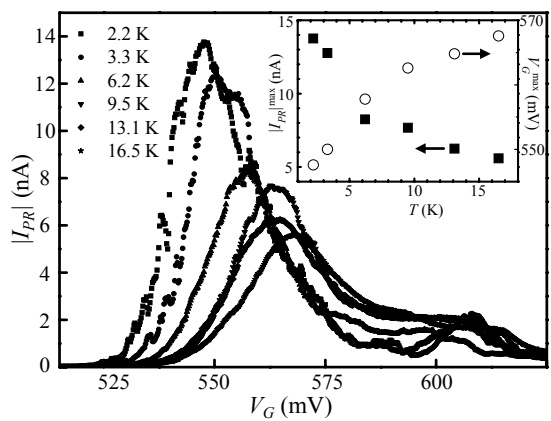

**FIGURE 2.** Photoresponse measurements at temperatures between $T = 2.2$ K and 16.5 K at a photon energy of 1.59 eV, $f_{CHOP} = 313$ Hz, $I \sim 300$ mW/cm$^2$ and $U_{SD} = -1.5$ mV. Inset: Temperature dependence of the maximum photoresponse $|I_{PR}|^{max}$ (filled squares, data taken from the main graph) and the corresponding gating voltage $V_{G,}^{max}$ (open circles).

## EXPERIMENTAL SECTION

The QWs are lithographically fabricated in a modulation-doped AlGaAs/GaAs heterostructure which contains a quantum well comprising a two-dimensional electron gas (2DEG). As depicted in Fig. 1(a)-(c), the QWs are defined by a lateral constriction within the 2DEG by a combination of e-beam lithography and chemical etching [11-14]. Further details of the heterostructure and the samples are described in [9,12]. All measurements are carried out in a helium continuous-flow cryostat at a vacuum of about 10$^{-5}$ mbar. Applying a voltage to the gate, the chemical potential of the subbands in the QWs is shifted with respect to the Fermi energies of the source/drain contacts. Typical for QWs, we detect conductance steps of $2e^2/h$ at low temperatures [Fig. 1(d)]. Illuminating the aperture of the devices with laser light, we measure the ac-photoresponse $|I_{PR}(f_{chop})| = |I^{ON}(f_{chop}) - I^{OFF}(f_{chop})|$ across the sample with the laser being in the "on" or "off" state. The signal is amplified by a current-voltage converter and detected with a lock-in amplifier utilizing the reference signal provided by the chopper frequency $f_{chop}$ [10]. We find that the photoresponse shows a maximum at the onset of each one-dimensional subband of the QW (see Fig 1(d) and [9]).

Fig. 2 (a) depicts photoconductance traces in the temperature range between 2 and 16.5 K. The maxima of the photoresponse (denoted as $|I_{PR}|^{max}$) shift to a more positive gate voltage for a higher temperature (inset of Fig. 2). We interpret the finding that for a higher temperature fewer holes are captured at the edges of the QWs, and in turn, a more positive gate voltage $V_G^{max}$ needs to be applied for a maximum photoresponse. At 17 K the thermal energy of about $k_BT \sim 2$ meV is comparable to the subband energy of the QWs of about 2-6 meV [9]. As a result, the recombination of the spatially separated electrons and holes is enhanced because of thermal activation, and the maximum photoresponse is decreased [Fig. 2].

## SUMMARY

In summary, we present photoresponse measurements on semiconductor quantum wire (QW) devices. By defining QWs with relatively large subband energies of 2-6 meV, we are able to perform experiments up to a temperature of about 17 K. We gratefully acknowledge financial support from BMBF via nanoQUIT, the DFG (Ho 3324/4), the Center for NanoScience (CeNS), the LMUexcellent program and the German excellence initiative via the "Nanosystems Initiative Munich (NIM)".